\documentclass{jfm}
\usepackage{epstopdf, epsfig}
\usepackage{amsmath,amssymb,amsfonts,bm}
\usepackage[euler]{textgreek}
\usepackage{rotating}
\usepackage{stmaryrd}
\usepackage{xcolor}
\usepackage{cancel}
\usepackage{graphicx}
\usepackage{subcaption}
\usepackage{wrapfig}
\usepackage{multicol}

\def\begineq{\begin{equation}}
\def\endeq{\end{equation}}

\def\begineqn{\begin{equation*}}
\def\endeqn{\end{equation*}}

\def\beginar{\begin{eqnarray}}
\def\endar{\end{eqnarray}}
\def\beginarn{\begin{eqnarray*}}
\def\endarn{\end{eqnarray*}}

\def\lb{\left ( }
\def\rb{\right ) }

\def\ub{\boldsymbol{u}}

\def\haz{\boldsymbol{\hat z}}
\def\hax{\boldsymbol{\hat x}}
\def\hay{\boldsymbol{\hat y}}

\newcommand\Beq{\begin{eqnarray}} 
\newcommand\Eeq{\end{eqnarray}}

\newcommand{\pd}[1]{\partial_{#1}}

\newcommand{\Ra}{\widetilde{Ra}}
\newcommand{\eps}{\varepsilon}

\shorttitle{Convection onset with Ekman pumping}
\shortauthor{S. Tro, I. Grooms and  K. Julien}

\title{Asymptotic approximations for convection onset with Ekman pumping at low wavenumbers}

\author{Sara Tro\aff{}\corresp{\email{sara.tro@colorado.edu}}, Ian Grooms\aff{}, Keith Julien\aff{}}

\affiliation{\aff{}Department of Applied Mathematics, University of Colorado, Boulder, CO 80309, USA}

\begin{document}

\maketitle

\vspace{1.5cc}

\begin{abstract} %250 words
\noindent  Ekman pumping is a phenomenon induced by no-slip boundary conditions in rotating fluids. In the context of Rayleigh-Bénard convection, Ekman pumping causes a significant change in the linear stability of the system compared to when it is not present (that is, stress-free). Motivated by numerical solutions to the marginal stability problem of the incompressible Navier-Stokes (iNSE) system, we seek analytical asymptotic solutions which describe the departure of the no-slip solution from the stress-free. The substitution of normal modes into a reduced asymptotic model yields a linear system for which we explore analytical solutions for various scalings of wavenumber. We find very good agreement between the analytical asymptotic solutions and the numerical solutions to the iNSE linear stability problem with no-slip boundary conditions.  
\end{abstract}
\vspace{0.95cc}
\parbox{30cc}{{\it Keywords: B\'enard Convection, Quasi-geostrophic flows,  Boundary layer stability  }%Fill maximum 5 key words
}
%\end{abstract}

\section{Introduction} \label{intro}

Rotating Rayleigh-Bénard convection (RRBC) is a relevant model in many geo- and astrophysical phenomena. In the case of rapid rotation, the bulk of the system is dominated by a balance between Coriolis and pressure gradient forces known as \textit{geostrophy}. In the ``upright" case, meaning the gravity and the rotation axis are aligned, stress-free boundary conditions are compatible with geostrophy and no boundary layer occurs at leading order. However, if no-slip boundary conditions are imposed, the dominant balance near the boundary is between viscous and Coriolis forces, in what is known as the Ekman layer. Conservation of mass in this layer gives rise to the physical phenomenon called Ekman pumping (suction). In computation, the Ekman layer is narrow and computationally expensive to resolve. Stress-free boundary conditions do not break down the geostrophy of the interior, and thus are often chosen for computational convenience. However, many studies have been done to explore how the Ekman layer does effect the flow. \cite{sC61} computes critical Rayleigh numbers for convection onset for both cases (stress-free and no-slip) and finds that these values differ for nonzero Ekman numbers ($E$). Asymptotic studies (\cite{pN65}, \cite{wH71}, \cite{zhang1993influence}) show that while nonzero Ekman numbers have a destabilizing effect, the difference is asymptotically small in the limit of rapid rotation, $E \to 0$. Studies of nonlinear effects find an increase in heat transport which does not vanish with $E$ (\cite{homsy1971}, \cite{sS2014}, \cite{jCKMSV16}, \cite{pachev2020rigorous}). In our recent publication, \cite{tro2024parameterized}, we discover that the marginal stability curve of the system with Ekman pumping departs significantly from the stress-free curve at small wavenumbers. We find that this effect does not vanish with $E\to 0$. In this paper, we describe that departure with analytical, asymptotic solutions.

\section{Formulation} \label{formulation}
 RRBC is governed by the incompressible Navier-Stokes equations (iNSE). Nondimensionalization by characteristic velocity scale $U$, horizontal length scale $\ell$, vertical depth scale $H$, advective timescale $\ell/U$, pressure scale $P$ and temperature difference $\Delta T$ results in nondimensional parameters including the Ekman number, $E$, and the reduced thermal Rayleigh number, $\Ra = Ra E^{4/3}$. We also have a Rossby number, $Ro$, which is small in the rapidly rotating limit, so we define $\eps$ such that
 \begin{equation}
 Ro=E^{1/3}\equiv\varepsilon\ll 1.
\end{equation}
We further select $U = \nu /\ell $ where $\nu$ is the kinematic viscosity, and $\ell/ H = \eps $. The resulting nondimensional iNSE system is given by
\begin{subequations}
\label{eq:NSE}
\beginar
\pd{t}\ub +\ub \cdot \nabla \ub +\frac{1}{\eps}\ \haz\times \ub + \frac{1}{\eps} \nabla p &=& \nabla^2 \ub +\frac{\Ra}{\sigma}
 \theta \haz,\\
\pd{t}\theta +\ub \cdot \nabla\theta -\eps\  \haz\cdot\ub &=& \frac{1}{\sigma}\nabla^2 \theta,\\
\nabla \cdot \ub &=& 0,
\endar
\end{subequations}
where $\ub$ and $p$ are the velocity and pressure fields, and $\theta$ is the temperature with the linear background profile removed, that is, $\theta = T +z-1$, where $T$ is the temperature field.  The parameter $\sigma =\nu/\kappa  $ is the Prandtl number, where $\kappa$ is the thermal diffusivity.

We consider a local $f$-plane approximation to the spherical shell, and focus on the case where gravity and the rotation axis are aligned. The unit vectors $\hax$, $\hay$, $\haz$, define the coordinate directions and point east, north, and radially. The components of $\ub$ are $(u,v,w)$ and are respectively in the $\hax$, $\hay$, $\haz$ directions.

The system (\ref{eq:NSE}) is accompanied by the boundary conditions 
\begin{equation}
    w = 0,\quad \theta = 0,\quad Z = 0,1
\end{equation}
 where $Z$ is the non-dimensional coordinate in the $\haz$ direction, and either 
 \begin{subequations}
     \beginar
        \mbox{no-slip:}& \quad (u,v) &= 0 \quad Z=0,1,\\
       \mbox{stress-free:}&\quad  \haz \cdot \nabla (u,v) &= 0\quad Z = 0,1.
     \endar
 \end{subequations}

Using asymptotic matching, the fluid variables of the iNSE may be decomposed into an interior (outer) geostrophic solution, a middle thermal wind layer, and boundary (inner) Ekman layer solution. For no-slip boundary conditions, the result for the interior (outer) solution plus the middle thermal wind layer is the composite quasi-geostrophic equations (CQG) given in \cite{jCKMSV16},
\begin{subequations}
\label{eq:CQGRBC}
    \beginar
\pd{t}\nabla_\perp^2\psi +J\left[\psi,\nabla_\perp^2\psi\right] -\pd{Z}w &=& \nabla_\perp^4 \psi,\\
 \pd{t} w +J\left[\psi,w\right]  +\pd{Z} \psi &=& \nabla_\perp^2 w +\frac{\Ra}{\sigma} \theta',\\
 \pd{t} \theta'  +J\left[\psi, \theta' \right] +w\lb \pd{Z}\overline{\Theta}-1\rb &=& 
  \frac{1}{\sigma}\nabla^2 \theta',
 \endar
\beginar \label{eqn:pumping_BC}
w = \delta \frac{\eps^{1/2}}{\sqrt{2}}\nabla_\perp^2 \psi,\quad \theta = 0,\quad\mbox{on}\  Z = 0,1,
\endar
\end{subequations}
where $\delta = 1$ at $Z=0$ and $\delta = -1$ at $Z=1$. Here, $\psi$ is the stream function such that $\pd{x}\psi = v$, $\pd{y}\psi = -u$, and $\theta = \overline{\Theta}+\eps \theta'$ has been split into a lateral mean (denoted by the overbar) and fluctuating component (denoted by the prime). The operators  $\nabla = (\pd{x},\pd{y},\eps \pd{Z})$, $\nabla_\perp =(\pd{x},\pd{y})$, and $J\left[\psi,\cdot \right]= \pd{x}\psi\pd{y} - \pd{y}\psi\pd{x}$. The boundary condition (\ref{eqn:pumping_BC}) on $w$ is known as the \textit{pumping} boundary condition since it encapsulates the pumping/suction effect of the Ekman boundary layer on the interior solution. Note that in this model, the vertical diffusion of $\theta'$ is retained due to the inclusion of the $\mathcal{O}(\eps) = \mathcal{O}\left(E^{1/3}\right)$ thermal wind layer. Without the vertical derivatives, the thermal boundary conditions cannot be satisfied. 

We explore the linear stability of both the rescaled iNSE (\ref{eq:NSE}) and the CQG model (\ref{eq:CQGRBC}) by substituting the normal mode ansatz
\begin{equation}
  \boldsymbol{v} =   \breve{\boldsymbol{v}}(Z)\exp\left( s t + i \mathbf{k}_\perp\cdot\mathbf{x}_\perp\right)
  \label{eq:normalmodes}
\end{equation}
for convective rolls, where we define the wavenumber $\mathbf{k}_\perp = (k_x,k_y)$. Inputting parameters $\eps$, $\sigma$, $k_\perp = |\mathbf{k}_\perp|$, and $s=0$ for marginal stability, the result is a differential equation eigenvalue problem for $\Ra$. The iNSE problem is then solved numerically (see \cite{tro2024parameterized} for the numerical formulation), and we explore asymptotic solutions to the CQG linear stability problem. While our following analysis is general for $\sigma = \mathcal{O}(1)$, note that by solving the problem with $s=0$, we are neglecting oscillatory convection for $\sigma<1$.

\begin{figure}
    \centering
    \includegraphics[width=0.88\linewidth]{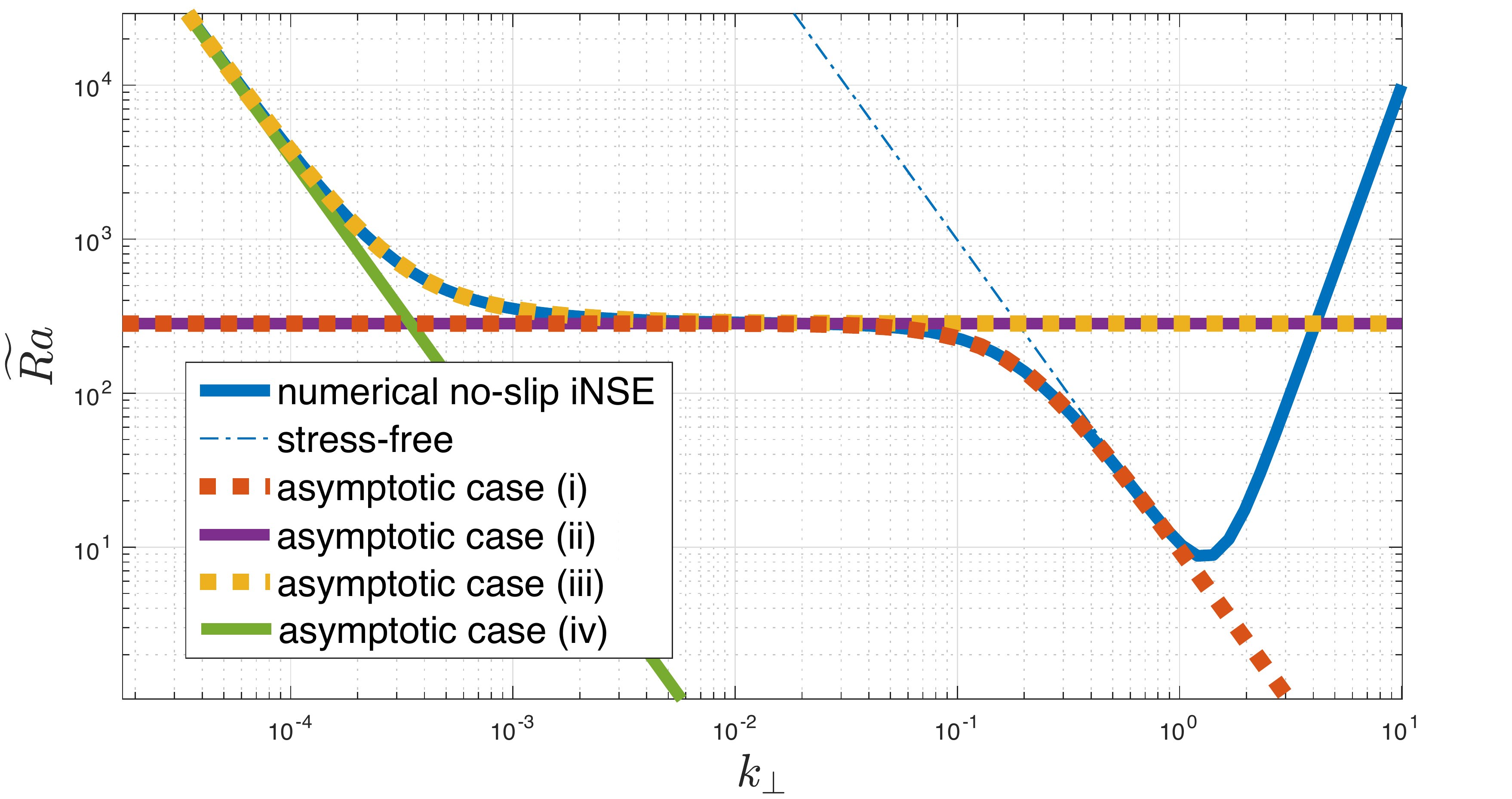}
    \caption{Marginal stability of the system in the $\Ra$-$k_\perp$ plane for  $\eps = 10^{-4}$ and $\sigma = 1$. The numerically computed curve for the iNSE system with Ekman pumping is shown in solid blue. The blue dash-dot curve is the stress-free case, $\Ra = k_\perp^4+n^2\pi^2/k_\perp^2$. The asymptotic approximations for cases \textit{(i)-(iv)} are in red, purple, yellow, and green respectively.}
    \label{fig:marginalcurves}
\end{figure}

Assuming $w_0\sim \psi_0\sim \theta_0 \sim k_\perp^2 \sim \Ra_0\sim 1$, where the subscript denotes the leading order term in an asymptotic expansion, the pumping boundary condition is subdominant at $\mathcal{O}(\eps^{1/2})$. This means that the boundary condition at leading order is $w_0 = 0$ at $Z = 0,1$. This yields the linear stability problem given in \cite{KJ98}, with solution 
\begin{subequations}
    \beginar
 w &= & A\sin(n\pi Z)+ \mathcal{O}(\eps^{1/2}) \\
 \psi &= & -A\frac{n\pi}{k_\perp^4}\cos(n\pi Z)+ \mathcal{O}(\eps^{1/2})\\
 \theta &= & A\frac{\sigma}{k_\perp^2}\sin(n\pi Z)+ \mathcal{O}(\eps^{1/2})
    \endar
\end{subequations}
and eigenvalue
\begin{equation}
    \Ra \sim  \frac{k_\perp^6+n^2\pi^2}{k_\perp^2},
    \label{eq:Ra_marginal}
\end{equation}
which is the same as found in \cite{sC61}.
The curve (\ref{eq:Ra_marginal}) in the $k_\perp$-$\Ra$ plane is the marginal stability curve, meaning that for parameter values above the curve, there is a positive growth rate $s$ and the system is unstable to perturbation. For stress-free boundary conditions in the rapidly rotating limit $\eps \to 0$, the marginal stability curve of the iNSE problem will converge to (\ref{eq:Ra_marginal}). However, in \cite{tro2024parameterized} we discovered that (\ref{eq:Ra_marginal}) does not describe the marginal stability of the no-slip problem at low wavenumbers $k_\perp$. This numerical result is reproduced in Figure \ref{fig:marginalcurves} by the solid blue line, contrasted with the stress-free result in blue dash-dot.

In this paper, we seek an analytical description of the marginal stability curve when Ekman pumping is present, shown numerically in \cite{tro2024parameterized} and plotted in Figure \ref{fig:marginalcurves}. The pumping boundary condition remains $\mathcal{O}(1)$ when $w\sim \eps^{1/2}k_\perp^2\psi \sim 1$, ($\psi \sim k_\perp^{-2}\eps^{-1/2}$). Under this scaling, we expand all dependent variables and $\Ra$ in half powers of $\eps$, and consider only the leading order ($\mathcal{O}(1)$) solution. Four cases occur: (i) $k_\perp^2\sim \eps^{1/2}$, (ii) $\eps^{1/2}\gtrsim k_\perp^2 \gtrsim \eps^2 $, (iii) $k_\perp^2 \sim \eps^2 $, and (iv) $ k_\perp^2 \lesssim \eps^2$, each of which is outlined in section \ref{sec:results} below, and each of which describes a section of the marginal stability curve.

\section{Results} \label{sec:results}
To ensure pumping remains $\mathcal{O}(1)$ as we explore $k_\perp^2 \lesssim \eps^{1/2}$, we begin by rescaling $\psi \to k_\perp^{-2}\eps^{-1/2}\check{\psi}$, where $\check{\psi}\sim 1$. Note that throughout the manuscript we use the inverted hat (check) to denote the rescaled quantities, including on their asymptotic expansions. 
Applying this scaling to the linear problem associated with (\ref{eq:CQGRBC}) (dropping the prime on the fluctuating temperature), yields 
\begin{subequations}
    \beginar
     -\pd{Z} w&=& k_\perp^2\eps^{-1/2} \check{\psi}\\
    \pd{Z}\check{\psi}&=& k_\perp^2 \eps^{1/2}\frac{\Ra}{\sigma}\theta - k_\perp^4\eps^{1/2} w\\
     - w &=& \frac{1}{\sigma}\lb \eps^2  \pd{Z}^2 -k_\perp^2 \rb\theta
    \endar
    \label{eq:scaledlinear}
\end{subequations}
with 
\begin{equation}
    w = -\frac{\delta}{\sqrt{2}} \check{\psi},\quad \theta =0,\quad  Z = 0,1.\ %\delta =  \pm1
\end{equation}
Under this scaling, the horizontal diffusion of $w$, the term $k_\perp^4 \eps^{1/2} w$, will always be less than $\mathcal{O}(\eps^{1/2})$ for the $k_\perp^2<1 $ cases considered below. We expand the eigenfunctions in powers of $\eps^{1/2}$, but only consider the leading order terms, 
\begin{equation}
    w = w_0 + \mathcal{O}(\eps^{1/2}), \quad \psi = \psi_0 +\mathcal{O}(\eps^{1/2}), \quad \theta = \theta_0 + \mathcal{O}(\eps^{1/2}).
\end{equation}
We also expand the eigenvalue $\Ra$ in the same way 
\begin{equation}
    \Ra = \Ra_0 + \mathcal{O}(\eps^{1/2}).
\end{equation}

\subsection{Case (i): $k_\perp^2\sim \eps^{1/2}$}
In this case vortex stretching is balanced by the horizontal diffusion of $\check{\psi}$ in (\ref{eq:scaledlinear}$a$). In (\ref{eq:scaledlinear}$c$), the horizontal diffusion of $\theta$ is dominant over the vertical, and to bring this into balance with $w$, we scale $\theta \to k_\perp^{-2}\check{\theta}$. Finally, we rescale $\widetilde{Ra}\to \eps^{-1/2}\widehat{Ra}$ such that buoyancy is in balance with the vertical gradient of $\check{\psi}$ in ($\ref{eq:scaledlinear} b$). This results in the leading order system 
\begin{subequations}
    \beginar
     -\pd{Z} w_0&=& k_\perp^2\eps^{-1/2} \check{\psi}_0\\
    \pd{Z}\check{\psi}_0 &=& \frac{\widehat{Ra}_0}{\sigma}\check{\theta}_0\\
     - w_0 &=& -\frac{\check{\theta}_0}{\sigma}
    \endar
    \label{eq:case1system}
\end{subequations}
with boundary conditions
\begin{equation}
    w_0 = -\frac{\delta}{\sqrt{2}} \check{\psi}_0, \quad  Z = 0,1,\ \delta =  \pm.
    \label{eq:BCs_notemp}
\end{equation}
This has solution
\begin{subequations}
    \beginar
w_0(Z) &=& A\lb \frac{\eps^{1/2}}{\sqrt{2}}\frac{\Ra_0^{1/2}}{k_\perp} \cos\lb Z \sqrt{k_\perp^2\Ra_0}\rb+ \sin\lb Z \sqrt{k_\perp^2\Ra_0}\rb \rb \\
    \psi_0(Z) &=& A \frac{\Ra_0^{1/2}}{k_\perp^3} \lb -\cos\lb Z \sqrt{k_\perp^2\Ra_0}\rb+\frac{\eps^{1/2}}{\sqrt{2}}\frac{\Ra_0^{1/2}}{k_\perp} \sin\lb Z \sqrt{k_\perp^2\Ra_0}\rb \rb 
    \endar
    \label{eq:case1efuns}
\end{subequations}
and $\theta_0 = \sigma k_\perp^{-2} w_0$. The eigenvalue $\Ra_0$ must satisfy 
\begin{equation}
    \frac{\sqrt{2\eps}\sqrt{k_\perp^2\Ra_0}}{k_\perp^2}\cos\lb \sqrt{k_\perp^2\Ra_0}\rb +\lb 1-\frac{\eps\Ra_0}{2k_\perp^2}\rb\sin\lb \sqrt{k_\perp^2\Ra_0}\rb=0.
    \label{eq:case1Racond}
\end{equation}
The numerically computed root to this equation for a range of $k_\perp^2$ is presented in Figure \ref{fig:marginalcurves} in the red dotted line. It agrees well with the underlying blue curve (the numerically-computed marginal curve for the iNSE) near $k_\perp \approx .1 = \eps^{1/4}$, as expected. If we consider the limit of (\ref{eq:case1Racond}) as $k_\perp^2$ becomes smaller than $\eps^{1/2}$, $\sin\lb \sqrt{k_\perp^2\Ra_0}\rb \sim \sqrt{k_\perp^2\Ra_0}$, and $\cos\lb \sqrt{k_\perp^2\Ra_0}\rb \sim 1$, we find that $\Ra_0 \to 2\sqrt{2/\eps}$, which is consistent with (\ref{eq:case2Ra}) below.

\begin{figure}
    \centering
    \includegraphics[width=\linewidth]{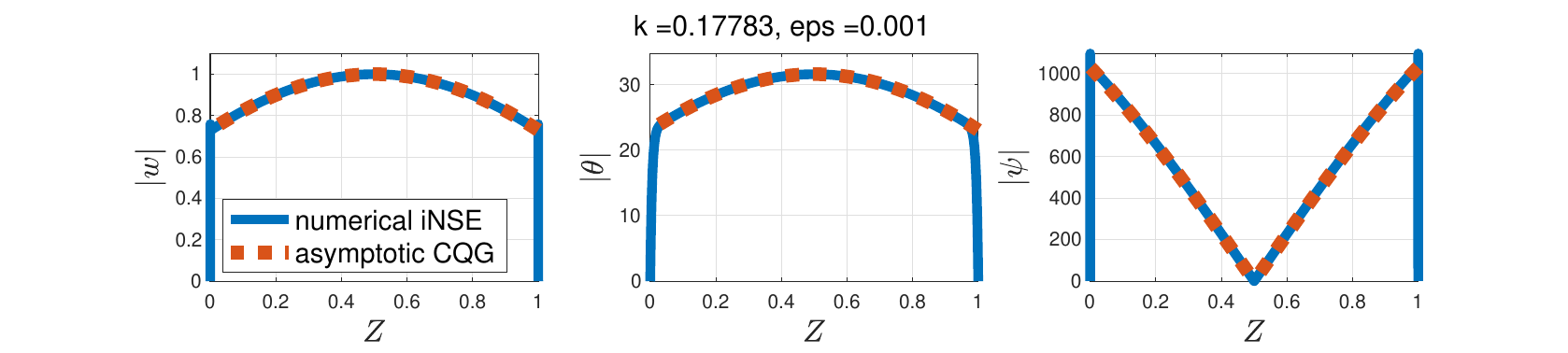}
    \caption{Eigenfunction profiles for case \textit{(i)}, $k_\perp^2\sim \eps^{1/2}$. We have selected as an example $\eps = 10^{-3}$ and $k_\perp = \eps^{1/4}$. Numerically computed eigenfunctions to the iNSE problem are in solid blue, and the analytical asymptotic approximations (\ref{eq:case1efuns}) are in dashed red. }
    \label{fig:case1efuns}
\end{figure}

Choosing in particular $k_\perp^2 = \eps^{1/2}$ and the corresponding $\Ra$ satisfying (\ref{eq:case1Racond}), examples of the vertical profiles (\ref{eq:case1efuns}) are plotted in Figure \ref{fig:case1efuns} in dotted red. The underlying blue curves are the eigenfunctions of the numerically computed iNSE marginal stability problem. We see very good agreement between the two curves in the interior of the domain. The analytical asymptotic expressions for $w$ and $\psi$ derived from the CQG model do not match the numerical solutions from the iNSE in the boundary layer because the CQG model parameterizes rather than resolves the Ekman boundary layer. The temperature boundary condition is not satisfied due to the subdominance of the vertical diffusion of $\theta_0$, but there is a boundary layer of size $\eps/k_\perp \sim \eps^{3/4}$ that allows $\theta_0 = 0$ to be satisfied.

\subsection{Case (ii): $\eps^{1/2}\gtrsim k_\perp^2 \gtrsim \eps^2 $}

\begin{figure}
    \centering
    \includegraphics[width=\linewidth]{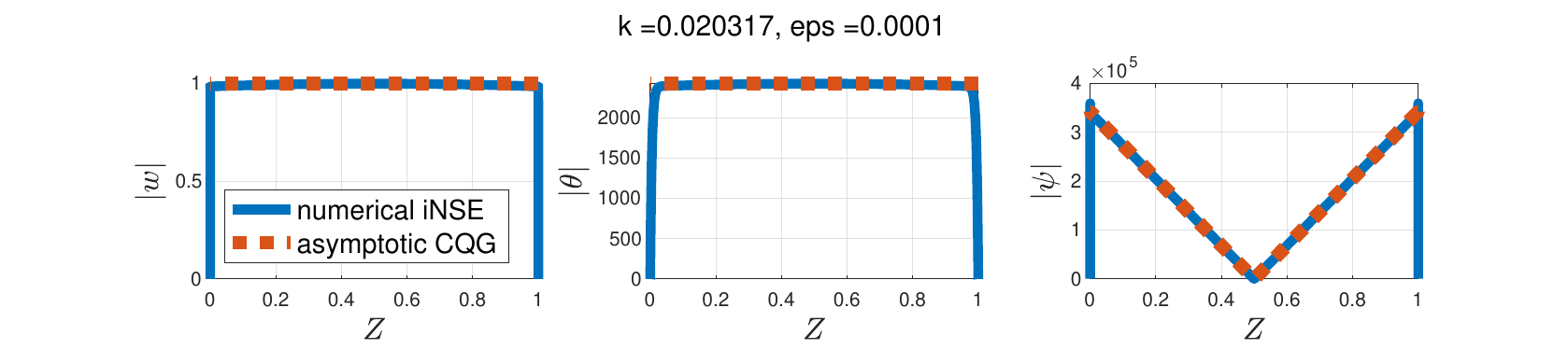}
    \caption{Eigenfunction profiles for case (ii), $\eps^{1/2}\gtrsim k_\perp^2 \gtrsim \eps^2 $, with $\eps = 10^{-4}$ and $k_\perp \approx .0203$. Numerically computed eigenfunctions to the iNSE problem are in solid blue, and the analytical asymptotic approximations (\ref{eq:case2efuns}) are in dashed red.}
    \label{fig:case2efuns}
\end{figure}

In this case, the horizontal diffusion of temperature is still dominant over the vertical diffusion, so we take $\theta \to k_\perp^{-2} \check{\theta}$, and again to retain the buoyancy term we take $\Ra \to \eps^{-1/2}\widehat{Ra}$. This yields the system
\begin{subequations}
    \beginar
    \pd{Z}w_0 &=& 0,\\
    \pd{Z}\check{\psi}_0 &=& \frac{\widehat{Ra}_0}{\sigma}\check{\theta}_0,\\
    w_0&=& \frac{\check{\theta}_0}{\sigma},
    \endar
    \label{eq:case2system}
\end{subequations}
with boundary conditions the same as above in (\ref{eq:BCs_notemp}). The horizontal diffusion of $\psi_0$ is now subdominant to vortex stretching, so $\pd{Z}w_0 = 0$, which gives a constant for $w_0$. The solution is
\begin{subequations}
    \beginar
w_0 &=&A,\\
\psi_0&=&A\ \frac{\Ra_0}{k_\perp^2}\lb Z-\frac{1}{2}\rb,\\
\theta_0&=& \frac{\sigma A}{k_\perp^2},
    \endar
    \label{eq:case2efuns}
\end{subequations}
with eigenvalue
\begin{equation}
    \widehat{Ra}_0 =2\sqrt{2}\quad \mbox{i.e.,}\quad \Ra_0 = \frac{2\sqrt{2}}{\eps^{1/2}}.
    \label{eq:case2Ra}
\end{equation}
The constant estimate for $\Ra$ given in (\ref{eq:case2Ra}) is shown in Figure \ref{fig:marginalcurves} in purple. We can see that it aligns well with the flat portion of the numerical iNSE curve. We also compare the eigenfunctions (\ref{eq:case2efuns}) to the numerical iNSE in Figure \ref{fig:case2efuns}. Again we see good agreement in the interior of the domain, but as the asymptotic result is designed to remove the boundary layer, they do not match in the Ekman boundary region. As in case \textit{(i)}, there is a thermal boundary layer of size $\eps/k_\perp$ which allows the temperature boundary condition to be satisfied. In this case, the boundary layer is still asymptotically thin
since $1 \gtrsim \eps/k_\perp\gtrsim \eps^{3/4}$.

\subsection{Case (iii): $k_\perp^2 \sim \eps^2 $} 

In this case, the vertical diffusion term of the temperature equation is now in balance with the horizontal, and we again use the scalings $\theta \to k_\perp^{-2}\check{\theta}$ and $\Ra \to \eps^{-1/2}\widehat{Ra}$. This gives us the leading order problem
\begin{subequations}
    \beginar
    \pd{Z}w_0&=& 0,\\
    \pd{Z}\check{\psi}_0&=& \frac{\widehat{Ra}_0}{\sigma}\check{\theta}_0, \\
     - w_0 &=& \frac{1}{\sigma} \lb \eps^2k_\perp^{-2}\partial_Z^2 - 1 \rb  \check{\theta}_0,
    \endar
    \label{eq:case3system}
\end{subequations}
with boundary conditions
\begin{equation}
    w_0 = -\frac{\delta}{\sqrt{2}} \check{\psi}_0, \quad \check{\theta}_0=0,\quad   Z = 0,1,\ \delta =  \pm 1.
    \label{eq:BCs_rescaled_wtemp}
\end{equation}
This has solution 
\begin{subequations}
    \beginar
    w_0 &=& A,\\
    \psi_0 &=&\frac{A\ \Ra_0}{k_\perp^2} \lb Z-\frac{1}{2} -\frac{\eps}{k_\perp}\frac{e^{k_\perp Z/\eps}}{1+e^{k_\perp/\eps}}+\frac{\eps}{k_\perp}\frac{e^{-k_\perp Z/\eps}}{1+e^{-k_\perp/\eps}} \rb, \\
       \theta_0 &=& \frac{\sigma A}{k_\perp^2}\lb 1-\frac{e^{k_\perp Z/\eps}}{1+e^{k_\perp/\eps}}-\frac{e^{-k_\perp Z/\eps}}{1+e^{-k_\perp/\eps}} \rb ,
    \endar 
    \label{eq:case3efuns}
\end{subequations}
with 
\begin{equation}
    \Ra_0 = \sqrt{\frac{2}{\eps}}  \lb \frac{1}{2} +\frac{\eps}{k_\perp}\frac{1}{1+e^{k_\perp/\eps}}-\frac{\eps}{k_\perp}\frac{1}{1+e^{-k_\perp/\eps}} \rb^{-1}.
    \label{eq:case3Ra}
\end{equation}
Here, the second-order derivative allows for the temperature boundary condition to be satisfied, and the thermal boundary layer width is $\mathcal{O}(1)$. 

In Figure \ref{fig:marginalcurves}, the $\Ra_0$ curve is plotted as a yellow dotted line. It agrees well with the underlying blue curve around the corner near $k_\perp\approx 10^{-4} = \eps$, as well as the region surrounding this value. If we consider (\ref{eq:case3Ra}) where $k_\perp \gg \eps $, it follows that $\Ra_0 \to 2\sqrt{2/\eps}$, which aligns with (\ref{eq:case2Ra}). We may also consider $k_\perp \ll \eps $, and using a few terms in the series $\displaystyle\frac{1}{1+e^{x}}\approx \frac{1}{2}+\frac{x}{2}-\frac{x^3}{48}$, we find that $\Ra_0 \sim 24\sqrt{2}\eps^{3/2}k_\perp^{-2}$. See below that this agrees with (\ref{eq:case4Ra}). 

\begin{figure}
    \centering
    \includegraphics[width=\linewidth]{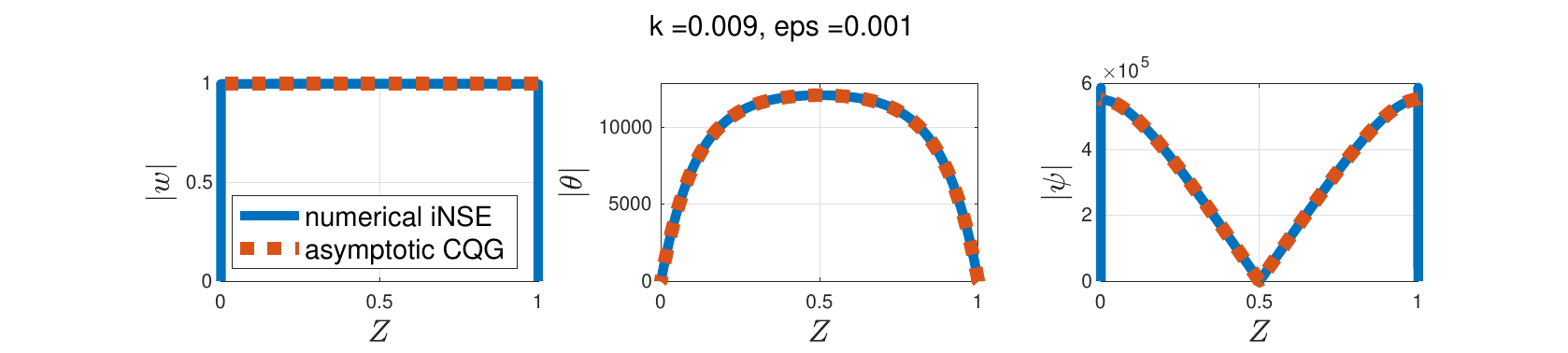}
    \caption{Eigenfunction profiles for case (iii), $k_\perp^2 \sim \eps^2 $, with $\eps = 10^{-3}$, $k_\perp = 9\cdot 10^{-3}$. Numerically computed eigenfunctions to the iNSE problem are in solid blue, and the analytical asymptotic approximations (\ref{eq:case3efuns}) are in dashed red.}
    \label{fig:case3efuns}
\end{figure}

We also plot the eigenfunctions in Figure \ref{fig:case3efuns}, where we again see good agreement with the numerical iNSE linear stability solution.

\subsection{Case (iv): $ k_\perp^2 \lesssim \eps^2$}

\begin{figure}
    \centering
    \includegraphics[width=\linewidth]{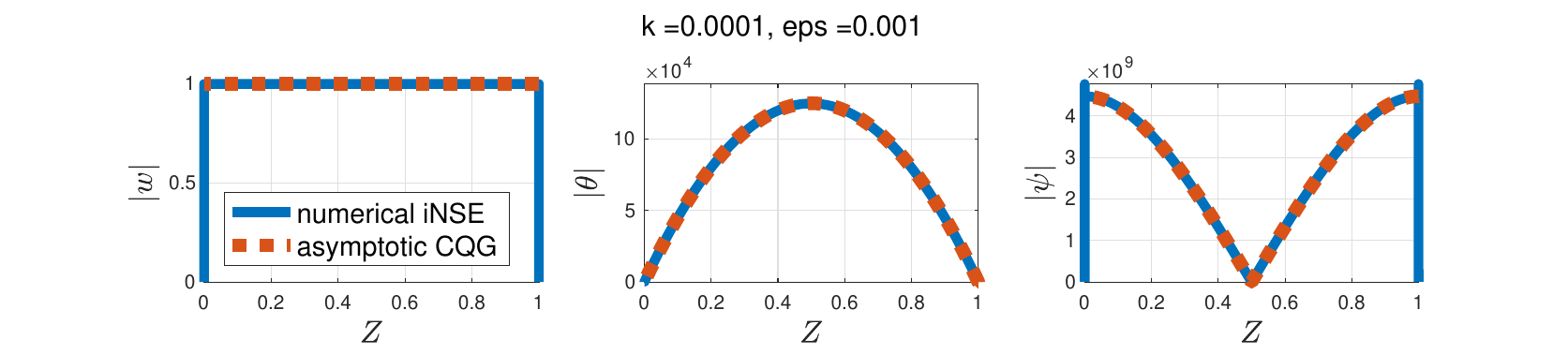}
    \caption{Eigenfunction profiles for case (iv), $ k_\perp^2 \lesssim \eps^2$, with $\eps = 10^{-3}$, $k_\perp = 10^{-4}$. Numerically computed eigenfunctions to the iNSE problem are in solid blue, and the analytical asymptotic approximations (\ref{eq:case4efuns}) are in dashed red.}
    \label{fig:case4efuns}
\end{figure}

In this case, we lose the horizontal diffusion term and the vertical diffusion of temperature is dominant. This implies the scalings $\theta \to \eps^{-2}\check{\theta}$ and $\Ra \to \eps^{3/2}k_\perp^{-2}\widehat{Ra}$, and yields the leading order system
\begin{subequations}
    \beginar
    \pd{Z}w_0&=& 0,\\
    \pd{Z}\check{\psi}_0&=& \frac{\widehat{Ra}_0}{\sigma}\check{\theta}_0, \\
     - w_0 &=& \frac{1}{\sigma} \partial_Z^2 \check{\theta}_0,
    \endar
    \label{eq:case4system}
\end{subequations}
with solution
\begin{subequations}
    \beginar
    w_0 &=& A,\\
    \psi_0 &=&A\ \frac{\Ra_0}{\eps^2} \lb -\frac{Z^3}{6}+\frac{Z^2}{4}-\frac{1}{24} \rb, \\
    \theta_0 &=& \frac{\sigma A}{2\eps^2 }Z\lb1-Z  \rb,
    \endar 
    \label{eq:case4efuns}
\end{subequations}
and eigenvalue 
\begin{equation}
    \widehat{Ra}_0 = 24\sqrt{2} \quad \mbox{i.e.,} \quad \Ra_0 = \frac{24\sqrt{2}\eps^{3/2}}{k_\perp^2}.
    \label{eq:case4Ra}
\end{equation} 
This curve is plotted in green in Figure \ref{fig:marginalcurves}, and is in good agreement with iNSE in the small $k_\perp$ limit. This eigenfunctions shown in Figure \ref{fig:case4efuns} are also in good agreement with iNSE.

\section{Discussion and Conclusions}

Through the four cases considered above, we have demonstrated that the linear stability of the rapidly rotating iNSE system with no-slip boundary conditions can be well represented by leading order asymptotic approximations. Linear stability theory for stress-free boundary conditions has been established for decades. In \cite{tro2024parameterized}, results of numerical computation shows a departure from that theory for no-slip boundary conditions. We have considered analytical asymptotic solutions to the linear stability problem for small wavenumbers $k_\perp^2 \lesssim 1$. The transition of $k_\perp^2$ to values smaller than $\eps^{1/2}$ leads to a loss of the lateral viscous term that normally balances vortex stretching, and this combined with the pumping boundary conditions leads to a constant, nonzero value for $w$ in the interior of the domain. The distinction between subsequent cases is whether the vertical and horizontal diffusion of temperature are in balance, or one is dominant over the other. At the smallest scales, for $k_\perp^2 < \eps^2$, vertical diffusion dominates. 

The condition that the pumping boundary conditions become $\mathcal{O}(1)$ is closely tied to the condition that $\pd{Z}w=0$. At large values of $k_\perp$, $k_\perp^4 \psi$, does not become subdominant to $\pd{Z}w$. Scaling the eigenfunctions such that $w\sim 1$ in the interior of the domain, by the vortex stretching balance, $1\sim w \sim k_\perp^4 \psi$. This implies $k_\perp^2\psi \sim k_\perp^{-2}$, making it impossible for pumping $\sim \eps^{1/2}k_\perp^2 \psi\sim \eps^{1/2}/k_\perp^2$ to become $\mathcal{O}(1)$ for large $k_\perp$. 

The CQG model is derived under the assumption that the aspect ratio of horizontal to vertical scales, $\ell/H$, is $\mathcal{O}(\eps)$ in the limit of $\eps \to 0$. In cases $(iii)$ and $(iv)$, when we consider $k_\perp \leq \eps$, we are violating that assumption and the validity of the CQG model. To justify this, we show in the appendix that if we begin from the iNSE system, we can derive the same systems of equations (\ref{eq:case3system}) and (\ref{eq:case4system}). In essence, any terms that could arise when vertical scales are not assumed smaller than horizontal scales remain subdominant due to the smallness of $k_\perp$.

From our results, we can estimate the value of $\Ra$ above which the departure from stress-free occurs. The stress-free marginal curve is shown in Figure \ref{fig:marginalcurves} in blue dash-dot. We can see for $\Ra$ below the purple line, $2\sqrt{2/\eps}$, the no-slip and stress free curves match up. However, if $\Ra$ is above $2\sqrt{2/\eps}$, we see a much wider range of wavenumbers contributing to the instability. From equation (\ref{eq:case4Ra}), we may estimate that wavenumbers
\begin{equation}
    \frac{24\sqrt{2}\eps^{3/2}}{\Ra} \leq k_\perp^2 \leq \Ra^{1/2},\quad \mbox{for}\quad \Ra > \frac{2\sqrt{2}}{\eps^{1/2}}
\end{equation}
produce modes with a positive growth rate. This is significantly different from the range for stress-free, 
\begin{equation}
    \frac{\pi^2}{\Ra} \leq k_\perp^2 \leq \Ra^{1/2},
\end{equation}
which does not change with $\eps$. Whether these modes affect direct numerical simulations of the fully nonlinear problem remains unexplored. 

\section*{Funding} This work was supported by the National Science Foundation (grant no. DMS-2308337).

\section*{Declaration of Interests} The authors report no conflict of interest.

\appendix
\section*{Appendix: Asymptotic justification for cases (iii) and (iv)}
\label{sec:appendix}
Beginning with the rescaled iNSE, equations (\ref{eq:NSE}), we consider the scalings of cases $(iii)$ and $(iv)$. Because $\psi$ is the streamfunction such that $\pd{x}\psi = v$, $\pd{y}\psi = -u$, and $\psi = p$, the rescaling $\psi \to k_\perp^{-2}\eps^{-1/2}\check{\psi}$ implies that 
\begin{equation}
    u\to k_\perp^{-2}\eps^{-1/2}\check{u}, \quad v\to k_\perp^{-2}\eps^{-1/2}\check{v}, \quad p \to k_\perp^{-2}\eps^{-1/2}\check{p}.
\end{equation}
For case $(iii)$, we take $\theta \to k_\perp^{-2}\check{\theta}$ and $\Ra \to \eps^{-1/2}\widehat{Ra}$ and for case $(iv)$, $\theta \to \eps^{-2}\check{\theta}$ and $\Ra \to \eps^{3/2}k_\perp^{-2} \widehat{Ra}$. Applying this scaling to the linear terms of (\ref{eq:NSE}$a$) with the normal mode ansatz (\ref{eq:normalmodes}) substituted in, and $s=0$, we get
\begin{subequations}
\beginar
-\check{v}+k_x\check{p} &=& \eps  \lb -k_\perp^2 +\eps^2 \pd{Z}^2 \rb \check{u} \\
\check{u}+k_y \check{p}&=&  \eps \lb -k_\perp^2 +\eps^2 \pd{Z}^2 \rb \check{v}   \\
0 & = & k_\perp \eps^{3/2} \lb  k_\perp^2 \eps^{1/2} \lb -k_\perp^2 +\eps^2 \pd{Z}^2 \rb w + \frac{\widehat{Ra} }{ \sigma} \check{\theta} - \pd{Z}\check{p}\rb ,
\endar
\label{eq:scaledathruc}
\end{subequations}
with (\ref{eq:NSE}$b$) becoming
\begin{subequations}
    \beginar
    0 =&   \displaystyle k_\perp \eps^{3/2}\lb w+\frac{1}{\sigma}\lb \eps^2k_\perp^{-2}\pd{Z}^2 -1\rb \check{\theta}\rb , \quad\quad &\mbox{for case } (iii)\\
    0 =&  \displaystyle k_\perp \eps^{3/2} \lb w + \frac{1}{\sigma}\pd{Z}^2 \check{\theta}\rb , \quad &\mbox{for case } (iv),
    \endar
    \label{eq:tempbycase}
\end{subequations}
and (\ref{eq:NSE}$c$) 
\begin{equation}
   \mathbf{k}_\perp\cdot \check{\ub}_\perp + k_\perp^2\eps^{3/2} \pd{Z}w =0,
   \label{eq:scaledincompress}
\end{equation}
where $\check{\ub}_\perp = \lb \check{u},\check{v} \rb$ and recall $\mathbf{k}_\perp = \lb k_x, k_y \rb $. We then expand the rescaled variables as follows:
\begin{subequations}
    \beginar
\check{u}  &=& \check{u}_0 +k_\perp \eps^{3/2}\check{u}_1 + ...\\
\check{v} &=& \check{v}_0 +k_\perp \eps^{3/2} \check{v}_1  +...\\
\check{p} &=& \check{p}_0 +k_\perp \eps^{3/2} \check{p}_1  +...\\
w &=& w_0 +...\\
\check{\theta} &=& \check{\theta}_0 +...
    \endar
\end{subequations}
Substituting in these expansions, The leading order system is 
\begin{subequations}
    \beginar 
-\check{v}_0 +k_x\check{p}_0 &=& 0,\\
\check{u}_0 +k_y\check{p}_0 &=& 0, \\
k_x \check{u}_0 +k_y \check{v}_0 &=& 0,
    \endar
    \label{eq:leadingorder}
\end{subequations}
which returns to us the streamfunction $\check{\psi}_0 = \check{p}_0$, with $k_x\check{\psi}_0 = \check{v}_0 $ and $k_y\check{\psi}_0 = -\check{u}_0$ as expected. 

To determine the next order problem, we must consider the scaling of $ -k_\perp^2 +\eps^2 \pd{Z}^2$ for cases $(iii)$ and $(iv)$, where $k_\perp^2 \leq \eps^2$:
\begin{equation}
    -k_\perp^2 +\eps^2 \pd{Z}^2 \sim \max\left\{ k_\perp^2,\ \eps^2 \right\} \sim \eps^2.
\end{equation}
Note also that $\mathbf{k}_\perp \sim k_\perp$. Then the $\mathcal{O}(k_\perp \eps^{3/2})$ terms are 
\begin{subequations}
    \beginar
-v_1 + k_x p_1 &=& 0 \\
u_1 +k_y p_1 & =&0 \\
k_x u_1 + k_y v_1 &=& -\pd{Z}w_0,
    \endar
    \label{eq:partoffirstordersys}
\end{subequations}
which has the solvability condition that 
\begin{equation}
    \pd{Z}w_0 = 0,
    \label{eq:solvability}
\end{equation}
and from (\ref{eq:tempbycase}), 
\begin{subequations}
    \beginar
    0 =&  w_0+\displaystyle\frac{1}{\sigma}\lb \eps^2k_\perp^{-2}\pd{Z}^2 -1\rb \check{\theta}_0, \quad &\mbox{for case } (iii)\\
    0 =& w_0 + \displaystyle\frac{1}{\sigma}\pd{Z}^2 \check{\theta}_0, \quad\quad\quad\quad &\mbox{for case } (iv).
    \endar
    \label{eq:tempappx}
\end{subequations}
and finally from (\ref{eq:scaledincompress}), 
\begin{equation}
    \pd{Z} \check{\psi}_0 = \frac{\widehat{Ra}_0}{\sigma}\check{\theta}_0.
    \label{eq:buoyanceappx}
\end{equation}
Then equations (\ref{eq:solvability}), (\ref{eq:tempappx}), and (\ref{eq:buoyanceappx}) match with (\ref{eq:case3system}) and (\ref{eq:case4system}).

\bibliographystyle{jfm}
\bibliography{ref}

\end{document}